\documentclass[twocolumn,showpacs,preprintnumbers,amsmath,amssymb]{revtex4}

\usepackage{longtable}
\usepackage{graphicx}
\usepackage{dcolumn}
\usepackage{bm}
\usepackage{graphicx, latexsym, amssymb, amsmath, color, multirow, mathrsfs, ifpdf}
\usepackage[section]{placeins}
  
\begin{document}

\title{ High Spin Spectroscopy and Shape Evolution  in $^{105}$Cd }
\author{ M. Kumar Raju$^{1,2}$}
\author{ D. Negi$^{1}$}
\author{ S. Muralithar$^1$}
\email{smuralithar@gmail.com}
\author{ R. P. Singh$^1$}
\author{ J. A. Sheikh$^{3,4}$}
\author{ G. H. Bhat$^{3}$}
\author{ R. Kumar$^1$}
\author{ Indu Bala$^1$}
\author{ T. Trivedi$^5$}
\author{ A. Dhal$^6$}
\author{ K. Rani$^1$}
\author{ R. Gurjar$^1$}
\author{ D. Singh$^7$}
\author{ R. Palit$^8$}
\author{ B. S. Naidu$^8$}
\author{S. Saha$^8$}
\author{ J. Sethi$^8$}
\author{R. Donthi$^8$}
\author{S. Jadhav$^{8}$}

\affiliation{$^1$Inter University Accelerator Centre, Aruna Asaf Ali Marg, New Delhi - 110067, India}
\affiliation{$^2$Department of Nuclear Physics, iThemba LABS, 7129, South Africa}
\affiliation{$^3$Department of Physics, University of Kashmir, Srinagar 190 006, India}
\affiliation{$^4$Department of Physics and Astronomy, University of Tennessee, Knoxville, TN 37996, USA  }
\affiliation{$^5$Guru Ghasidas Vishwavidyalaya, Bilaspur-495009, India}
\affiliation{$^6$Dept.of Particle and Astrophysics,Weizmann Institute of Science, Rehovot 76100, Israel}
\affiliation{$^7$Central University of Jharkhand,  Ranchi - 835 205, India} 
\affiliation{$^8$Tata Institute of Fundamental Research, Mumbai - 400005, India}



\date{\today}
             
\begin{abstract}
High spin states in $^{105}$Cd were studied  using $^{92}$Mo($^{16}$O, 2pn)$^{105}$Cd reaction at an incident beam energy of 75 MeV.    The level scheme of $^{105}$Cd has been observed up to J$^{\pi}$ = (47/2$^-$) and excitation energy $\sim$ 10.8 MeV with the addition of 30 new $\gamma$-transitions to the previous work. Spin and parity for most of the reported levels are assigned from the DCO ratios and linear polarization measurements.  The microscopic origin of the investigated band structures is discussed in the context of triaxial projected shell model (TPSM).  The energies of observed positive and negative parity  bands agree with the predictions of the TPSM by considering triaxial deformation for the observed excited band structures.  The shape evolution with increasing angular momentum is explained in the framework of  Cranked Shell Model (CSM) and the Total Routhian Surface (TRS) calculations.

\end{abstract}

\maketitle
\section{Introduction}
Study of nuclei in mass A $\sim$ 105 region close to the N = Z = 50 shell closures is interesting due to existence of various structural effects with increasing proton and neutron numbers.  The availability of limited number of valance nucleons with respect to doubly magic $^{100}$Sn core, makes this region, an ideal testing ground for the observation of various nuclear phenomena  such as  band termination, shears mechanism, antimagnetic rotation and shape evolution from collective to non-collective structures or vice versa \cite{103Cd, 104Cd,mag, anti, vib-rot}.  These structures are attributed to various coupling and competition between collective and single particle degrees of freedom.  

 The nuclei in this mass region show small deformation ($\beta_2 \approx 0.13$) \cite{beta12, beta22} at low spins, evolving into collective structures with increasing angular momentum.  This behavior is observed particularly in lighter odd-A Cd isotopes.  Since these nuclei lie in transitional region with two proton holes in high-$\Omega$ 1g$_{9/2}$ orbital and several neutron particles in low-$\Omega$ g$_{7/2}$, d$_{5/2}$ and h$_{11/2}$ orbitals, the competing rotational-vibrational structures are built on the configurations involving these orbitals.   For example, systematic study of $^{103-107}$Cd \cite{103Cd, D.Jerrestam, 107Cd} isotopes reveals that the yrast negative parity band structures are developed on vibrational excitation at low spins, which evolved into  rotational structures that terminate at higher spins.  These rotational structures are interpreted due to the alignment of $\nu$h$_{11/2}$ pair. Whereas the positive parity bands are interpreted as based on configurations involving $\nu$g$_{
7/2}$ and $\nu$d$_{5/2}$ orbitals.  Theoretically, these nuclei are predicted to have rapid  shape transitions as a function of rotational frequency due to the occupation probability of valance quasiparticles in deformation driving high-j, low-$\Omega$ orbitals. For example, such a shape transition is evident in $^{103}$Cd \cite{103Cd} based on total routhian surface calculations (TRS), which depicts that,  the prolate shape of this nucleus persists up to spin J$^\pi$=39/2$^-$ and changes to an oblate shape at higher spins.  In case of  $^{107}$Cd \cite{beta12}, the shape of the nucleus, which is quite $\gamma$-soft at low spin, evolves into a near prolate shape at higher spins.  Thus, the triaxial deformation parameter-$\gamma$ has strong influence over the shape of nuclei and it would be interesting to investigate such an effect in $^{105}$Cd.   

The low spin sates in $^{105}$Cd  have been reported previously by several groups \cite{alpha,16O}.   First identification of excited negative parity structures in $^{105}$Cd have been  done by Regan \emph{et al.,} \cite{Regan} using fusion evaporation reaction and reported the level structure up to J$^\pi$ = 47/2$^-$.  Later, the positive parity level structures were established by Jerrestam \emph{et al.,} \cite{D.Jerrestam} with several modifications made in negative parity sequences. The observed band structures were discussed based on the spin diabatic surface calculations. Recently, the lifetimes of the yrast negative parity states above spin 23/2 $\hbar$ have been measured using Doppler shift attenuation method \cite{105Cd-AMR}. The extracted $\it{B(E2)}$ values show decrease with increasing spin, whereas the dynamic moment of inertia remains constant at $\sim$ 30 $\hbar^2 MeV^{-1}$. These are signatures of anti-magnetic rotational phenomenon in this band.  Although, the yrast negative parity bands in 
$^{105}$Cd  have been reported up to high spins, the high spin state information for positive and non-yrast negative parity states in $^{105}$Cd is limited.  Therefore, detailed investigation  of these structures  in $^{105}$Cd to higher spins has been the subject of interest in the present work.  In addition, the study of shape evolution in this nucleus  enable us to understand the competing interplay between single particle and collective degrees of freedom in this region. The deduced band structures
have been analyzed using the cranked shell model (CSM) and triaxial projected shell model (TPSM)
approaches. 

The present article is organized as follows. In Section \ref{exp}, the experimental methods and data analysis procedure adopted in this work are briefly discussed. The experimental results and level scheme information are presented in Section \ref{results}.  Section \ref{discussion} includes discussion on the level structures based on the cranked shell model calculations and the shape evolution in $^{105}$Cd  in the light of TRS and triaxial projected shell model (TPSM) calculations. Finally, a brief summary is given in Section \ref{summary}. Preliminary  results of this work were reported in Ref.\cite{DAE}.

\section{Experimental Details and Data Analysis}\label{exp}
In the present experiment, high spin states in $^{105}$Cd  were populated using the fusion evaporation reaction $^{92}$Mo($^{16}$O, 2pn)$^{105}$Cd at an incident beam energy of 75 MeV.  Beam of $^{16}$O ions with current of $\sim$ 1 pnA was delivered by the 14UD Pelletron accelerator at Tata Institute of Fundamental Research (TIFR), Mumbai. The target used in the experiment was of 1 mg/cm$^2$ thickness on 10 mg/cm$^2$ Au backing.  The de-exciting $\gamma$-rays from reaction products were detected by the Indian National Gamma Array (INGA) \cite{INGA} facility at TIFR.  During this experiment, INGA setup comprised of fifteen Compton suppressed clover Ge detectors, out of which, four were placed at 90$^\circ$, two at 40$^\circ$, two at 65$^\circ$, two at 115$^\circ$, two at 140$^\circ$, and the remaining three were at 157$^\circ$ with respect to the beam direction.  The clover detectors were kept at a distance of 25 cm from the target.  The beam energy of 75 MeV was chosen based on  experimental excitation 
function (relative yield as a function of beam energy) and PACE-4 \cite{pace}   statistical model calculations.  At this beam energy, the cross section for $^{105}$Cd is dominant and is competing with other reaction product $^{105}$In.  The relative photo peak efficiency of INGA array and energy calibration were performed using $^{152}$Eu and $^{133}$Ba standard radioactive sources.   

The time stamped data were collected in list mode  using a Digital Data AcQuisition (DDAQ) system based on XIA Pixie-16 modules \cite{DDAQ} and trigger was set when at least two detectors were fired in coincidence.   A total of more than two billion $\gamma-\gamma$  and higher fold coincidence events were  recorded. The measured coincidence events were sorted in to $\gamma-\gamma$ matrices using the sorting program MARCOS (Multi Parameter Coincidence Search) developed at TIFR.    The data were analyzed in offline  using RADWARE \cite{radware} and CANDLE \cite{candle} analysis programs.  The level scheme of $^{105}$Cd has been constructed based on the $\gamma$-$\gamma$ coincidence relationships, intensity arguments, 
with the assumption that the intensity of the $\gamma$-transitions decreases monotonically as we go higher in cascade.  Relative intensities of $\gamma$-transitions were determined using the gated projections from the E$_\gamma$ - E$_{\gamma}$ symmetric matrix.

The multipolarity of the $\gamma$-transitions were assigned using the observed coincidence angular correlations \cite{dco}. For this purpose,  angle dependent matrices were  constructed by taking energies of the $\gamma$-transitions from all the detector at forward or backward angle on one axis and the coincidence $\gamma$-transitions from the rest of the detectors at  90$^\circ$ on other axis.  The experimental DCO ratio was defined as the intensity ({\it I}) of a measured $\gamma$-ray transition at  157$^\circ$ when gated on a reference $\gamma$-ray at 90$^\circ$, divided by the intensity of a measured $\gamma$-ray transition at 90$^\circ$ when gated on a reference $\gamma$-ray at  157$^\circ$ (where the reference $\gamma$-ray is of known multipolarity) and is given by the relation

\begin{equation}
R_{DCO} = \dfrac{I_{\gamma_1} \;\;at \;\; 157^{\circ} \;\;gated   \;\;by\;\; \gamma_2 \;\;at\;\; 90^\circ}{I_{\gamma_1} \;\;at\;\; 90^\circ \;\;gated   \;\;by\;\; \gamma_2 \;\;at\;\; 157^\circ} 
\end{equation}

If the gating transitions is of stretched quadrupole multipolarity then this ratio is $\sim$ 1 for stretched quadrupole transitions and $\sim$ 0.5 for stretched dipole ones.  If the gating transition is of stretched dipole multipolarity then this ratio is $\sim$ 2 for stretched quadrupole, and is $\sim$ 1 for  stretched dipoles.  For mixed dipole/quadrupole transitions R$_{DCO}$, depends on the value of mixing ratio. 

\begin{figure}

\includegraphics[scale=0.34,angle=270]{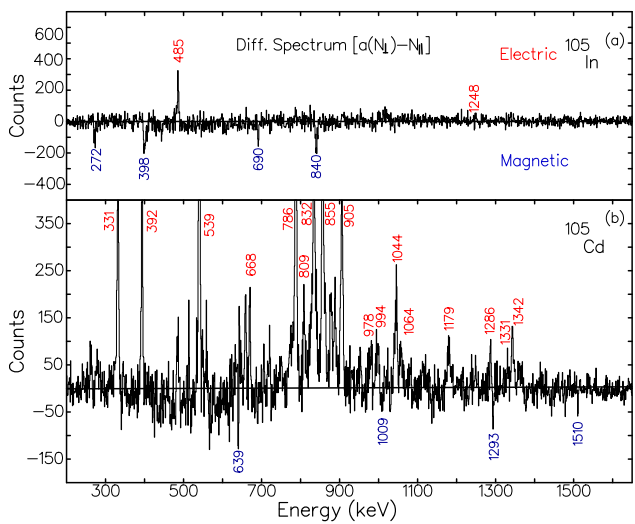}

\caption{\label{pol}(Color online)Difference spectrum with a normalization constant between perpendicular and parallel scattering events of $\gamma$-ray transitions in $^{105}$In (upper panel) and $^{105}$Cd (lower panel) distinguish the electric and magnetic nature of $\gamma$-rays. $^{105}$In difference spectrum is taken as reference.}
\end{figure}

\begin{figure}
\vspace{0.0cm}
\includegraphics[scale=0.34,angle=270]{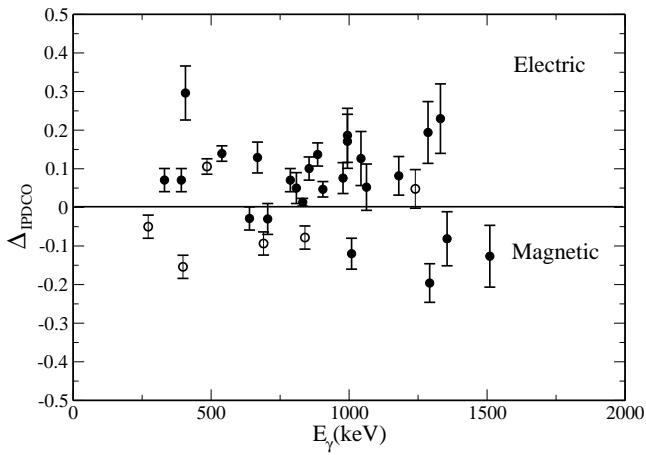}

\caption{\label{ipdco}Experimental polarization asymmetry (IPDCO) values are plotted for transitions in $^{105}$Cd (closed circles). Strong know transitions in $^{105}$In (open cirecles) are also included for reference.}
\end{figure}

Multipolarity assignments were further corroborated by measuring the linear polarization of the $\gamma$-rays using the integrated polarization directional correlation from oriented nuclei (IPDCO) method \cite{pol1, pol2}. The HpGe clover detector were used as Compton polarimeters to determine the electromagnetic character of the $\gamma$-transitions. Two polarization matrices were constructed from the data corresponded to energy recorded in all detectors on one axis, while the other axis corresponded to the energy scattered in a perpendicular or parallel segment of the Clover detectors at 90$^\circ$ with respect to the beam axis. The number of parallel (N$_\parallel$) and perpendicular (N$_\perp$) scattering events for a given $\gamma$-ray were obtained from projection spectra by gating on particular transitions in $^{105}$Cd. The experimental polarization asymmetry (IPDCO) ratio was deduced using the relation 

\begin{equation}
\bigtriangleup_{IPDCO} = \dfrac{a(E_\gamma)N_\perp-N_\parallel}{a(E_\gamma)N_\perp+N_\parallel} 
\end{equation}

Where N$_\perp$ and N$_\parallel$ are the number of counts of $\gamma$-transitions scattered perpendicular and parallel to the reaction plane.  a(E$_\gamma$) is a correction factor defined as ratio of N$_\parallel$/N$_\perp$ which is a measure of asymmetry between parallel and perpendicular scattering events with in the crystals of the clover detector. The value of a(E$_\gamma$) was found to be 0.99 obtained from the decay data of known radioactive sources $^{133}$Ba and $^{152}$Eu. The polarization asymmetry (IPDCO) ratio is expected to be positive for electric transitions and negative for magnetic transitions. A near zero value is indicative of possible admixture of electric and magnetic nature. Figure \ref{pol} illustrate the difference spectrum between perpendicular and parallel scatterers with correction factor. It shows the electric and magnetic nature of the transitions in $^{105}$Cd and $^{105}$In \cite{105In}. We included the difference spectrum of $^{105}$In to illustrate the electric and magnetic nature of transitions, as $^{105}$In is one of the strongest channels populated in the reaction.  The experimental polarization asymmetry (IPDCO) values obtained for several of the $\gamma$-transitions in $^{105}$Cd and $^{105}$In are shown in Fig. \ref{ipdco} and the $^{105}$Cd IPDCO values are summarized in Table~\ref{DCO}.

\section{Results}\label{results}
\begin{figure*}

\includegraphics[scale=1.0,angle=270]{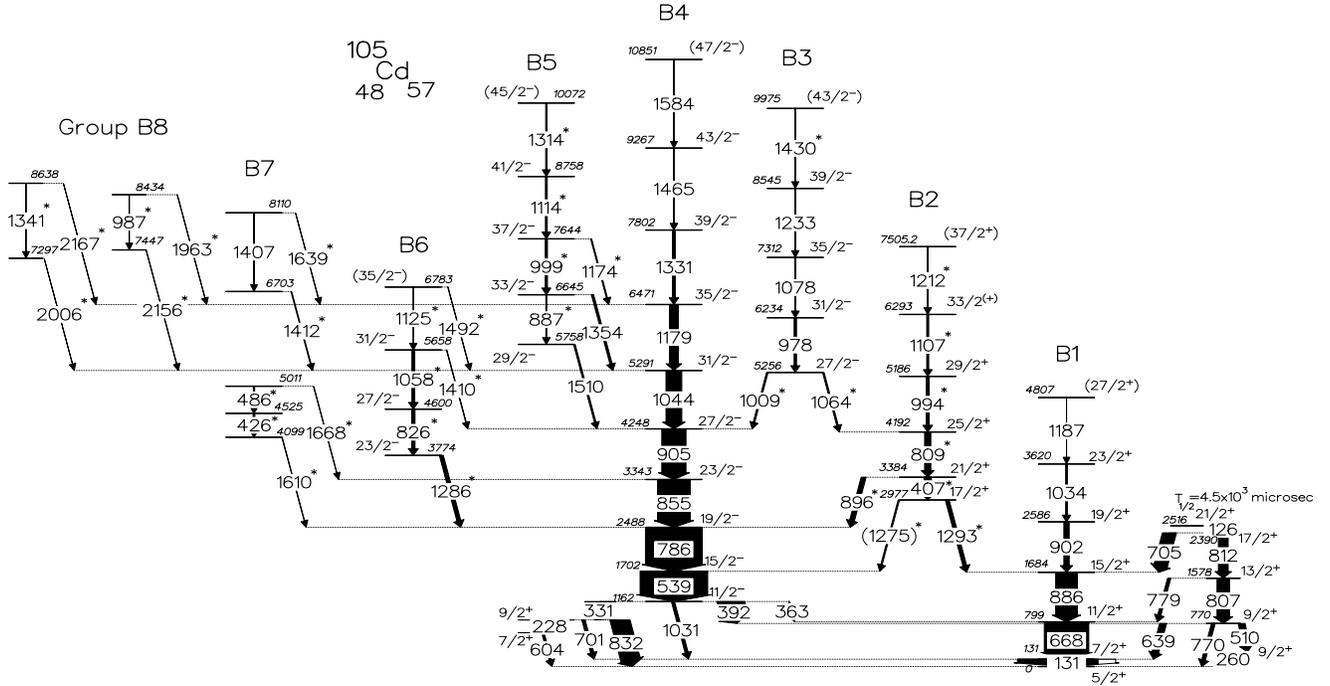}

\caption{\label{gls}Partial level scheme of $^{105}$Cd as obtained in the present work. Bands are labeled as B1 to B7 for reference in the text. The newly identified transitions are marked with (*) and the widths of the $\gamma$-transitions are approximately proportional to their intensities.}
\end{figure*}


\begin{figure}
\includegraphics[scale=0.37,angle=270]{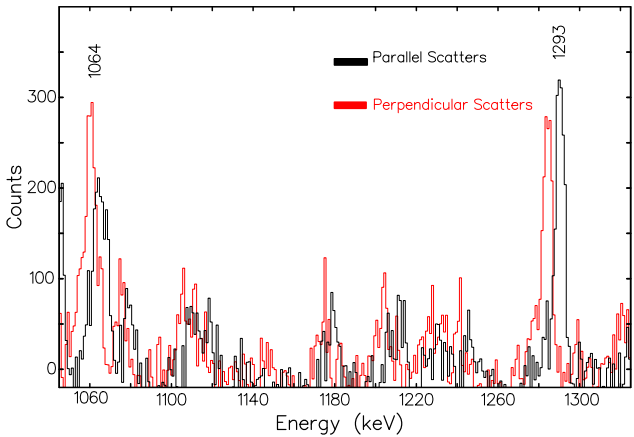}

\caption{\label{1293}(Color online)Shows the 407 + 809 keV gated projection spectrum showing the 1293 keV linking transition from band B2 to band B1 and the 1064 keV linking transition from band B3 to band B2.  The parallel components are slightly shifted for better visualization. Higher counts for 1064 keV transition in the perpendicular scattered spectrum indicates its  electric nature whereas higher counts for 1293 keV transition in the parallel scattered spectrum indicates its magnetic nature.}
\end{figure}

\begin{figure}
\includegraphics[scale=0.35,angle=270]{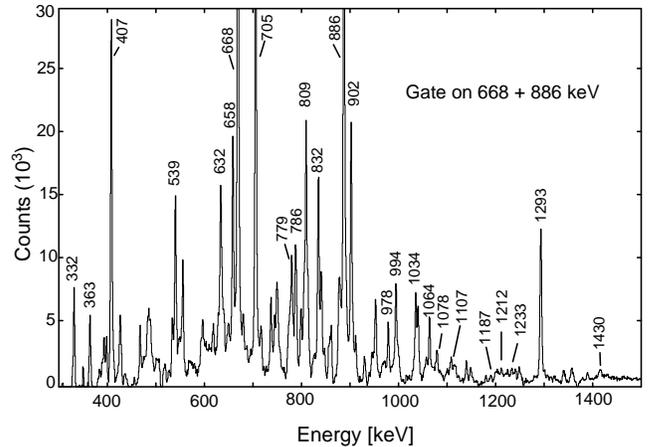}

\caption{\label{pp}Summed $\gamma$-$\gamma$ coincidence spectrum of  $^{105}$Cd with gates on 668, 886 keV transitions in the yrast positive parity sequence.}
\end{figure}
\subsection{Level Scheme of $^{105}$Cd}
The partial level scheme of $^{105}$Cd as obtained in the present work is shown in Fig. \ref{gls}.    The present work verified and confirmed the energy levels in the  previously known level schemes \cite{D.Jerrestam, Regan, 105Cd-AMR} with certain changes and additions.  A total of 30 new $\gamma$-ray transitions were identified and placed in the level scheme based on the coincidence sum energy relationships and intensity flow. The bands in the level scheme have been labeled namely B1 to B7.  A new positive parity band B2 is established in the present work  up to J$^\pi$ = (37/2$^+$). Three new negative parity bands are also identified and labeled  as band B3, B5 and B6 in the level scheme.    The spins and parities for the known low-lying states were adopted from the previous work \cite{D.Jerrestam}, and these values along with the measured DCO and polarization values were used as inputs for  spin and parity assignments to the newly identified states.  The $\gamma$-ray energies, measured relative intensities, directional correlation orientation ratios (DCO), polarization asymmetry($\bigtriangleup$), 
multipolarities of the observed $\gamma$-transitions,  and spins-parities of the levels belonging to $^{105}$Cd are summarized in Table~\ref{DCO}. Brief description of the observed band structures shown in level scheme are given below.


\begin{figure*}
\includegraphics[scale=0.55,angle=270]{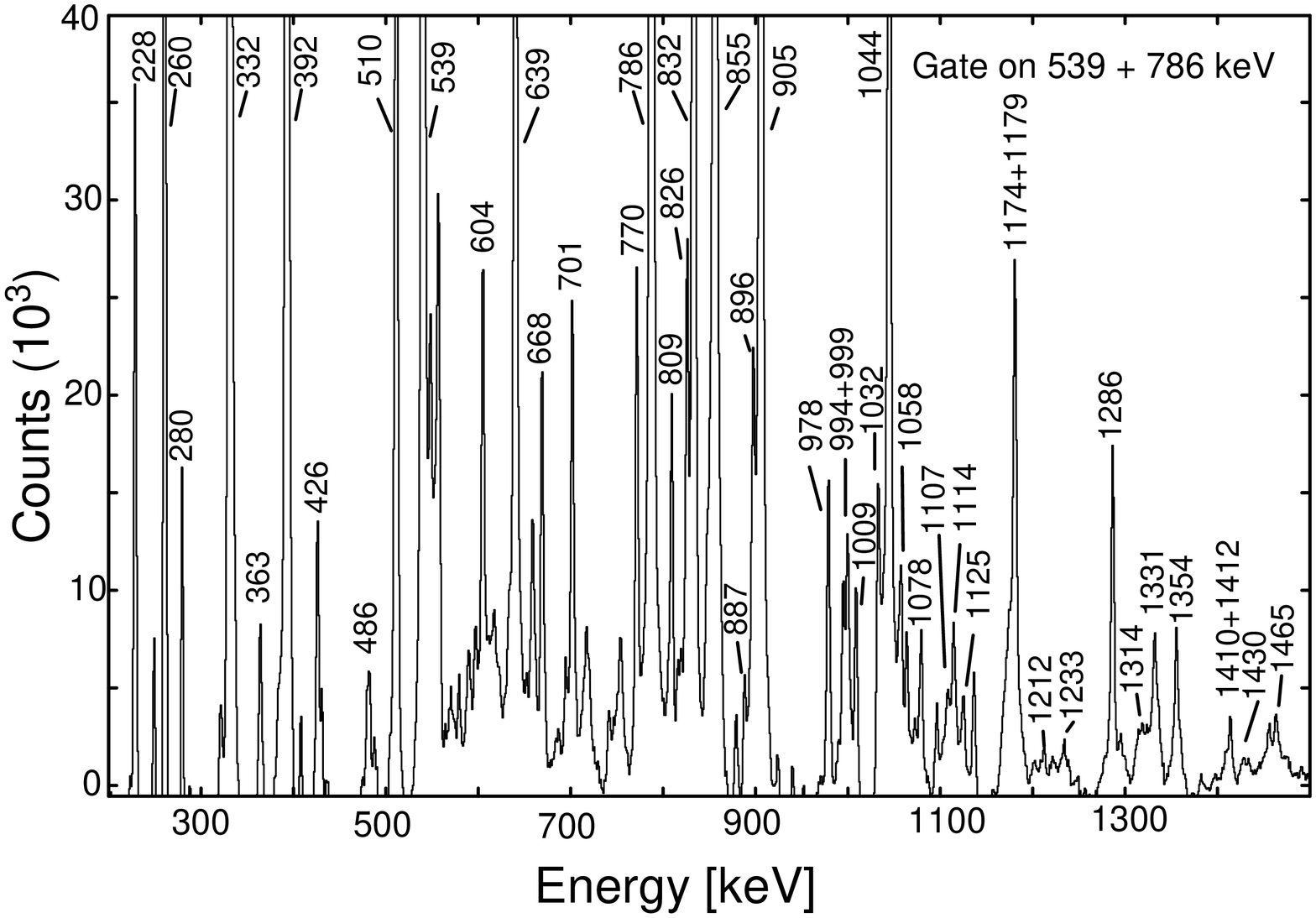}
\vspace{0.3cm}
\caption{\label{np}Summed $\gamma$-$\gamma$ coincidence spectrum of  $^{105}$Cd with gates on 539, 786 keV transitions in the yrast negative parity sequence of band B4.}
\end{figure*}

\subsubsection{\textbf{Positive parity bands}}\label{ppb}
The yrast positive parity band B1 was reported by Jerrestam \emph{et al.,} \cite{D.Jerrestam} up to  spin J$^\pi$ = (27/2$^+$) and this band was reported to decay from J$^\pi$ = (27/2$^+$)  to  (23/2$^+$) through 1293 keV $\gamma$-ray.  In this work, we could confirm the states in band B1 up to spin 23/2 $\hbar$.  The 1293 keV transition has been found to be in anti-coincidence with 1034, 902 keV decaying transitions and in coincidence with low spin 886, 668 and 131 keV $\gamma$-ray transitions. Therefore,   the 1293 keV transition is placed above 15/2$^+$.  The measured R$\it_{DCO}$ and IPDCO values for the 1293 keV indicate magnetic dipole character (previous work assigned E2 nature) which suggests the level spin to be 17/2$^+$ at an excitation energy of 2977 keV. Figure \ref{1293} is a representative polarization spectrum gated on 407+809 keV showing the overlap of  parallel (N$_\parallel$) and perpendicular (a(E$_\gamma$)N$_\perp$) scattering events observed in the 90$^\circ$ clover detectors. Higher counts for 1293 keV in parallel scattered spectrum indicates its magnetic nature while the reverse condition suggests the electric character for 1064 keV transition. A $\gamma$-ray of energy 1187 keV  is observed in coincidence with remaining transitions in band B1. Hence, this transition is placed above  the 23/2$^+$ state and the level at 4808 keV assigned a tentative spin J$^\pi$ = (27/2$^+$).  A series 
of $\gamma$-ray transitions 770, 807, 812, and 126 keV are decaying to the level with spin J$^\pi$ = 5/2$^+$ of band B1.  Though these transitions are quite intense, we could not observe transitions above 2516 keV level as this state is as an isomer with T$_{1/2}$ = 4.5 $\mu$s \cite{isomer}. 

A new positive parity band B2 is identified for the first time in this work (shown in Fig. \ref{gls}) consisting of a series of transitions 407, 809, 994, 1107, and 1212 keV, which are  placed above the level with spin  J$^\pi$ = 17/2$^{+}$.  A coincidence sum spectrum gated on 668, and 886 keV of band B1 displaying the transitions in band B1 and B2 is shown in  Fig.~\ref{pp}.  The measured R$\it_{DCO}$ and IPDCO values for the 407, 809, 994 and R$\it_{DCO}$ for 1107 keV transitions indicate their E2 nature. This suggests spins 21/2$^{+}$, 25/2$^{+}$, 29/2$^{+}$ and 33/2$^{(+)}$ for levels in this band.  We could not measure the DCO for 1212 keV $\gamma$-ray due to low statistics.  However, based on the coincidence and intensity arguments, this transition is placed above 33/2$^{(+)}$, which extends the band B2 up to a tentative spin J$^\pi$ = (37/2$^+$).  The positive parity for this band B2 is further supported based on the systamatics.  The isotones of $^{105}$Cd in lower mass region such as $^{103}$Pd \cite{103Pd} and 
$^{101}$Ru \cite{101Ru} have similar positive parity band structures, in which the positive parity yrast bands are established over ground state positive parity band.  These excited positive parity bands also have linking transitions to the yrast negative parity band. Band B2 is also one of such band developed over ground state and also have linking transitions to the yrast negative parity band.   Along with these new observations,  a  $\gamma$-ray transition of energy 1064 keV is observed  in coincidence with the transitions  below 25/2$^{+}$  level and  with transitions above 27/2$^-$ level in negative parity band B3. The measured DCO and IPDCO values of 1064 keV is consistent with electric dipole nature (see figure \ref{1293}) which confirm its placement between band B2 and band B3.  The placement of this transition is crucial, which causes the change of spins for all the states in band  B3.  Details about band B3 transitions will be discussed in the subsequent sections. The cross talk between transitions in band B2 and states below 19/2$^-
$ of negative parity band B4 is confirmed with the observation of 896 keV transition, and a tentative 1275 keV linking transition.  The low spin positive  parity states are confirmed in this work and fit well with the previous works \cite{Regan, D.Jerrestam}.

\setlength{\tabcolsep}{18pt}
\begin{table*}
\centering
\caption{Transition energy (E$_\gamma$), relative Intensity (I$_\gamma$), DCO ratio (R$_{DCO}$), Polarization asymmetry (IPDCO), multipolarity of the transition (Electric/magnetic  or D:Dipole/Q:Quadrupole), and initial and final spin and parity between which  γ-transitions placed in the level scheme of 105Cd, are listed.  The 539 keV ($\Delta$J = 2) transition in band B4 is used as gating transition for DCO measurements and the relative intensities are obtained with respect to the 539 keV transition by assuming its intensity as 100$\%$.  Errors are given in parentheses for I$_\gamma$, R$_{DCO}$ and IPDCO.  \label{DCO}}

\begin{tabular}{lllllll}
\hline\hline
 \emph{$E_{\gamma}$} & \emph{$I_{\gamma}$} & \emph{$R_{DCO}$}  & \emph{$\bigtriangleup_{IPDCO}$} &\emph{Multipolarity of} &\emph{$J_i^{\pi}$} &\emph{$J_f^{\pi}$}  \\
(keV)   &  (Rel.)       &               &    & \emph{transition}&  \\
\hline\hline  
126 & 14.5(7) & 0.97(3) &  & Q  & 21/2$^{+}$ &  17/2$^{+}$\\

131 & 97(8) & 0.56(3) &  &  M1\textsuperscript{a}   & 7/2$^{+}$ &  5/2$^{+}$\\

228 & 4.6(11) & 0.55(4) &  & D & 9/2$_2^{+}$ &  7/2$_1^{+}$\\

260 & 12.2(10) & 0.61(7) &  &  M1+E2\textsuperscript{a} & 9/2$_1^{+}$ &  5/2$^{+}$\\

331 & 44.5(3) & 1.29(10) & 0.07(3) & E1 & 11/2$^{-}$ & 9/2$_2^{+}$\\

363 & 1.6(3) & 0.61(7) &  & D & 11/2$^{-}$ & 11/2$^+$\\

392 & 40.2(4) & 0.55(3) & 0.07(3) & E1   & 11/2$^{-}$ & 9/2$^{+}$\\

407 & 10.8(3) & 0.97(6) & 0.02(7)  & E2 & 21/2$^{+}$) & 17/2$^{+}$\\

426 & 2.7(3) &  &  &   &  \\

486 & 2.0(7) &  &  &   &  \\

510 & 14.4(6) & 1.08(9) &  &  D & 9/2$^+$ & 9/2$_1^+$\\

539 & 100  &  & 0.13(2) &  E2  & 15/2$^-$ & 11/2$^-$\\

604 & 3.5(5) & 0.62(14) & -0.004(8)  & M1+E2\textsuperscript{a}  & 7/2$_1^+$ & 5/2$^+$  \\

639 & 20.4(6) & 0.52(3)&  -0.03(3) & M1+E2\textsuperscript{a} & 9/2$^{+}$ & 7/2$^+$\\

668 & 66.4(12) & 0.98(4) & 0.12(3) &  E2 & 11/2$^+$ & 7/2$^+$\\

701 & 4.8(4) & 0.65(11)&  &   M1+E2\textsuperscript{a}  & 9/2$_2^+$ & 7/2$^+$\\

705 & 18.3(6) & 0.59(7) &  -0.03(4)  & M1 & 17/2$^+$ &  15/2$^+$\\

770 & 12.5(7) & 1.11(5) & &  E2\textsuperscript{a} & 9/2$^+$ & 5/2$^+$\\

779 & 4.7(4) & 0.63(7) &  & D  & 13/2$^+$ &  11/2$^+$\\

786 & 84.5(7) & 0.98(3) & 0.07(3)  & E2 & 19/2$^-$ & 15/2$^-$\\

807 & 20.1(11) & 0.99(3) & 0.04(4)  & E2   &13/2$^{+}$ &  9/2$^{+}$\\

809 & 9.3(11) & 0.97(7) & 0.05(4)  & E2 & 25/2$^{+}$ &  21/2$^{+}$\\

812 & 16.5(7) & 1.19(11) &  & Q & 17/2$^{+}$ & 13/2$^{+}$\\

826 & 5.0(4) & 0.91(6) &  & Q & 27/2$_1^{-}$ & 23/2$_1^{-}$\\

832 & 31.5(6) & 0.97(6) & 0.09(4) & E2 & 9/2$^{+}$ & 5/2$^{+}$\\

855 & 48.6(15) & 0.95(5)& 0.10(3) &  E2& 23/2$^{-}$ & 19/2$^-$\\

886 & 32.7(8) & 0.96(6)& 0.13(3) &  E2 & 15/2$^{+}$ & 11/2$^+$\\

887 & 2.1(4) &  &  & (Q) & 33/2$^{-}$ & 29/2$^-$\\

896 & 7.1(8) & 0.49(7) &  & E1 & 21/2$^{+}$ &  19/2$^{-}$\\

902 & 12.0(4) & 0.97(8) & 0.08(2) & E2 & 19/2$^+$ & 15/2$^+$\\

905 & 36.0(6) & 0.96(5)& 0.04(2) & E2 & 27/2$^-$ & 23/2$^-$\\

978 & 3.5(4) & 0.96(12)& 0.07(4) & E2 & 31/2$^{-}$ & 27/2$^-$\\

987 & $<$ 1 & &  &  & & \\

994 & 4.6(7) & 0.98(2) & 0.17(7)  & E2 &  29/2$^{+}$ & 25/2$^{+}$\\

999 & 3.3(4) & 0.92(9) & &  Q & 37/2$^{-}$ & 33/2$^-$\\

1009 & 2.2(4) & 0.79(10) & -0.12(4) & M1  & 27/2$^-$ & 27/2$^-$\\

1032 & 4.6(4) & - & - &  & 11/2$^-$ & 7/2$^+$\\

1034 & 2.5(4) & 0.95(9) &  & Q & 23/2$^{+}$ & 19/2$^+$\\

1044 & 23.5(4) & 0.98(6)& 0.12(7) & E2  & 31/2$^{-}$ & 27/2$^-$\\

1058 & 4.23(5) & 0.97(9) &  & (Q) & 31/2$_2^{-}$ & 27/2$_1^{-}$\\

1064 & 2.4(4) & 0.65(9) & 0.05(6) & E1 & 27/2$_1^{-}$ & 25/2$^{+}$\\

1078 & 1.9(4) & 1.12(10)& & Q  & 35/2$_1^{-}$ & 31/2$_1^+$\\

1107 & 1.9(3) & 1.14(11) & &  Q & 33/2$^{(+)}$ &  29/2$^{+}$\\

1114 & 2.4(3) & 1.15(11) &  & Q & 41/2$^{-}$ &  37/2$^{-}$\\

1125 & 1.5(4) &  &  & (Q) & (35/2$_2^{-}$) & 31/2$_2^-$\\

1174 & 1.5(4) &  &  & (D) & 37/2$^{-}$ & 35/2$^-$\\

1179 & 13.2(6) & 1.05(11)  & 0.08(5) & E2 & 35/2$^{-}$ & 31/2$^-$\\

1187 & 0.5(4) &  &  & (Q) & (27/2$^{+}$) & 23/2$^+$\\

1212 & 1.1(4) &  &  &(Q)  & (37/2$^{+}$) & 33/2$^{(+)}$\\

1233 & 1.2(4) & 0.95(11) &  & Q & 39/2$_1^{-}$ & 35/2$_1^-$\\

1275 & 1.2(4) &   &   & D & 17/2$^{+}$ & $15/2^-$\\

1286 & 5.8(7) & 0.99(5)  & 0.19(8) & E2 & 23/2$_1^{-}$ & $19/2^-$\\

1293 & 5.6(5) & 0.57(4)  &  -0.19(5) & M1 & 17/2$^{+}$ & 15/2$^+$\\

\hline\hline
\multicolumn{5}{l}{\textsuperscript{a}\footnotesize{Adopted from NNDC}}\\
\end{tabular}

\end{table*}

\setcounter{table}{0}
\begin{table*}
\caption{(Continued...)}
\centering
\begin{tabular}{lllllll}
\hline\hline
 \emph{$E_{\gamma}$} & \emph{$I_{\gamma}$} & \emph{$R_{DCO}$}  &\emph{$\bigtriangleup_{IPDCO}$}  & \emph{Multipolarity of} &\emph{$J_i^{\pi}$} &\emph{$J_f^{\pi}$} \\
(keV)   &  (Rel.)       &               &   &  transition   &  \\
\hline\hline 

1314 & 2.0(4) & - &  &  (Q) & (45/2$^{-}$) & 41/2$^-$\\

1331 & 4.1(6) & 0.97(12)& 0.22(9) & E2 & 39/2$^{-}$ & 35/2$^-$\\

1341 &  &  &  &  &  \\

1354 & 3.3(4) & 0.67(9)  & -0.07(7) & M1 & 33/2$^{-}$ & 31/2$^-$\\

1407 & 1.2(7) &  &  &  &  &  \\

1410 & $<$ 1&  &  & (Q) & 31/2$_2^{(-)}$ & 27/2$^-$\\

1412 & $<$ 1&  &  & &  & 31/2$^-$\\

1430 & 1.1(4) &  &  & & (43/2$_1^{-}$) & 39/2$_1^-$\\

1465 & 1.5(4) & 1.12(13) & &  Q & 43/2$^{-}$ & 39/2$^-$\\

1492 & $<$ 1  &  & &  (Q) & (35/2$_2^{-}$) & 31/2$^-$\\

1510 & 2.4(5) & 0.48(11)& -0.02(8) & M1 &  29/2$^{-}$ & 27/2$^-$\\

1584 & $<$ 1  &  &  & (Q) & (47/2$^{-}$) & 43/2$^-$\\

1610 & $<$ 1 &  &  & &  &  \\

1639 & $<$ 1 &  &  & &  & 35/2$^-$\\

1668 & $<$ 1 &  &  & & & 23/2$^-$ \\

1963 & $<$ 1 &  &  & &  & 35/2$^-$\\

2006 & $<$ 1 &  &  & &  & 31/2$^-$\\

2156 & $<$ 1 &  &  & &  & 31/2$^-$\\

2167 & $<$ 1 &  &  & &  & 35/2$^-$\\ 

\hline\hline
\multicolumn{5}{l}{\textsuperscript{a}\footnotesize{Adopted from NNDC}}\\

\end{tabular}

\end{table*}

\begin{figure*}
\begin{center}
\includegraphics[scale=0.55,angle=270]{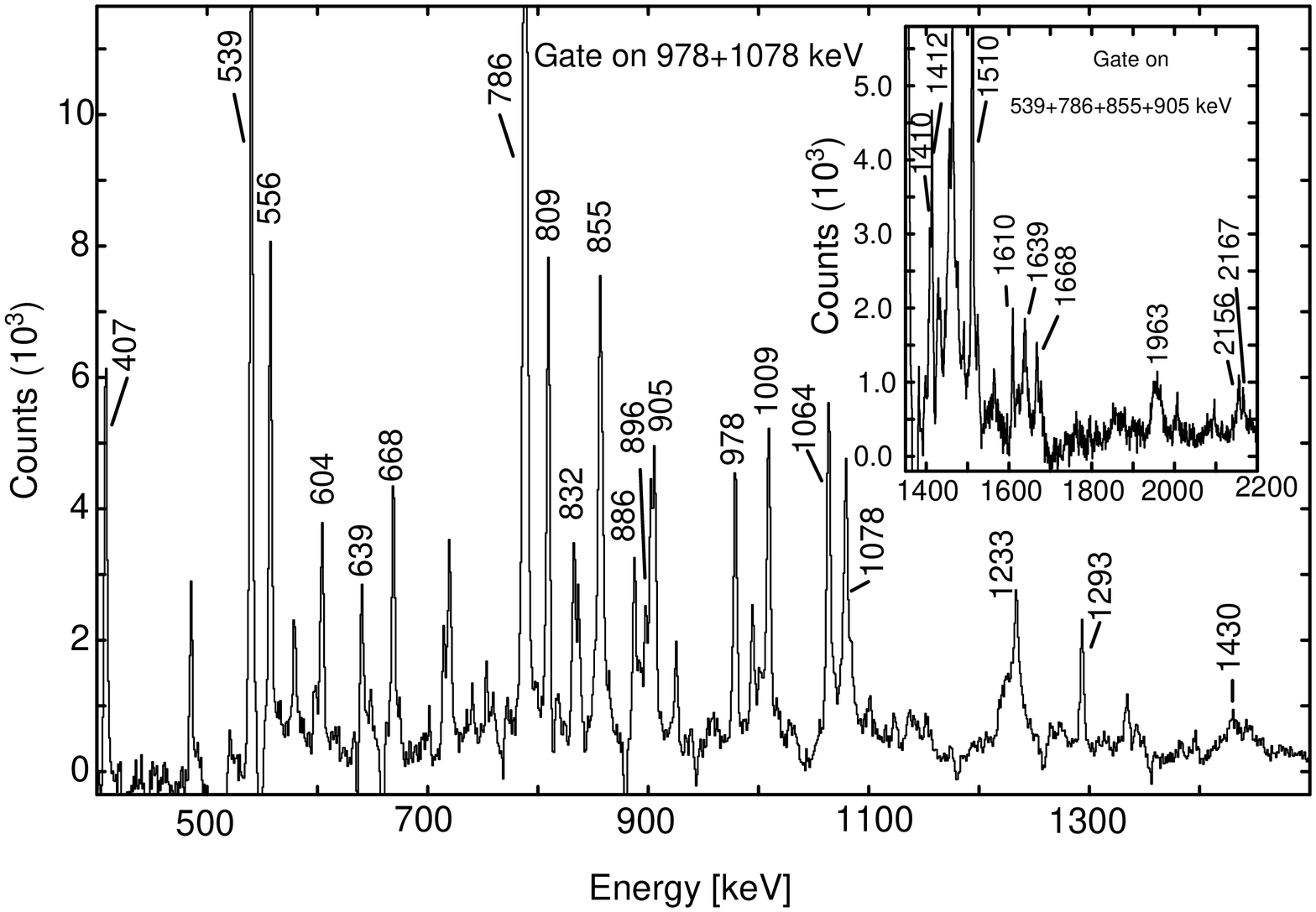}

\caption{\label{sum}Summed $\gamma$-$\gamma$ coincidence spectrum of  $^{105}$Cd with gates on 978, 1078 keV transitions in band B3. Inset illustrate the high energy  transitions above 1400 keV with gate on 539, 786, 855, and 905 keV.}
\end{center}
\end{figure*}

\subsubsection{\textbf{Negative parity bands}}
The proposed level scheme in the present work consists of four negative parity bands, namely band B3, B4, B5, and B6 (shown in Fig. \ref{gls}).  Band B4 is the strongly populated band, proposed to have $\nu h_{11/2}$ band head configuration \cite{Regan}.  In the present work, this band is observed up to a spin J$^\pi$ = (47/2$^-$) and excitation energy $\sim$ 10.8 MeV and agree well with the reported levels in the previous works \cite{Regan, D.Jerrestam}.  A  $\gamma-\gamma$ coincidence sum spectrum gated on 539, and 786 keV transitions of the yrast negative parity band B4 is shown in Fig. \ref{np}. This figure shows the newly identified transitions in bands B3, B5, B6, and B7 along with the known transitions in band B4.   Band B3 is reported up to a spin (41/2$^-$)  in Ref.\cite{D.Jerrestam}. We could extend this band up to  J$^\pi$ =  (43/2$^-$) and excitation energy 9.9 MeV with the addition of 1430 keV transition and made several modifications in spin parity assignments of this band.   The negative 
parity for this band B3 was assigned based on IPDCO and from previous works \cite{Regan, D.Jerrestam, 105Cd-AMR}.  A $\gamma$-ray of energy 1009 keV is placed as a connecting transition between band B3 and band B4.  This transition is observed  in coincidence with all the transition in band B3 and transitions below 27/2$^-$ in band B4. The measured DCO ratio for the 1009 keV $\gamma$-ray indicate $\Delta$J = 0 dipole nature (monopoles not allowed) suggesting  a spin J$^\pi$ = 27/2$^-$ to the level at 5256 keV.  This placement is further supported by the 1064 keV ($\Delta$J = 1) connecting transition between band B2 and B3, discussed in the previous section \ref{ppb}. A typical $\gamma-\gamma$ coincidence sum spectrum gated on 978, and 1078 keV transitions is shown in Fig. \ref{sum}.  This spectrum shows evidence for 1064, and 1009 keV linking transitions between bands B3 to  B2 and B3 to B4 respectively.   The 978 keV $\gamma$-ray in band B3 was previously reported  as magnetic dipole transition, but in the present work, we have assigned a $\Delta$J = 2 
electric character 
based on the DCO and IPDCO measurements. So the inclusion of the 1009, and 1064 keV transitions and assignment of $\Delta$J =2 nature to  978 keV transition changes the spin of each state in band B3 with respect to previous works \cite{Regan, D.Jerrestam, 105Cd-AMR}.  

\begin{figure}
\begin{center}

\includegraphics[scale=0.33,angle=270]{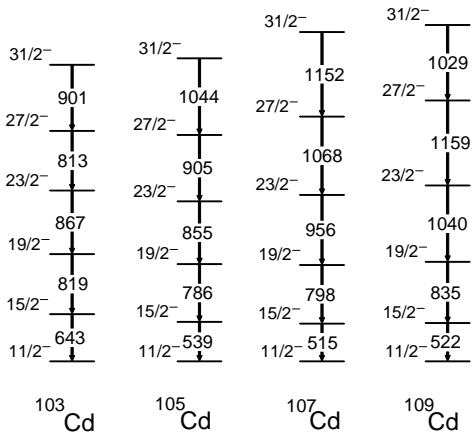}

\caption{\label{compare}Comparison of low spin energy levels  based on J$^\pi$ = 11/2$^-$ band of odd-A Cd isotopes.}
\end{center}
\end{figure}

Band B5 is a new negative parity sequence identified in the present work, built on the 29/2$^-$  level at 5758 keV and was extended by four new transitions
at energies of  887, 999, 1114, and 1314 keV.  This band  decays to the negative parity states in band B4 via linking transitions of energies 1510, 1354, and 1174 keV.  The observed R$\it_{DCO}$ of the 999, 1114 keV $\gamma$-ray transitions are consistent with $\Delta$J =2 nature, suggesting  J$^\pi$ = 37/2$^-$, and 41/2$^-$, respectively. This is further confirmed by the dipole magnetic nature of the linking transitions of energies 1510, 1354, and 1174 keV between  bands B5 and B4. Thus negative parity is assigned for the transitions in band B5.   The present work also established a negative  parity band B6 built on the 23/2$^{-}$ state at 3774 keV, decaying to the yrast  negative parity band B4 via a newly observed 1286 keV transition.  This band is observed to  spin 35/2 $\hbar$ with three $\gamma$-transitions of energies 826, 1058, 1125 keV.  The observed R$\it_{DCO}$ of  826, 1058, and decaying transition 1286 keV are consistent with $\Delta$J =2 nature.  We could not measure DCO ratio for the 1125 keV transition due to low statistics. However, its placement as part of band B6 can be confirmed via other observed linking transitions 1410 and 1492 keV decaying to band B4. The parity of the band is assigned tentatively based on the observed DCO ratios and IPDCO of 1286 keV linking transition between band B3 and B6.   In addition, a new sequence B7 is identified, which decays to the negative parity states 19/2$^-$, 23/2$^-$  in band B4 via linking transitions 1610, 1668 keV and consists of in-band transitions 426 and 486 keV and placed according to the coincidence and intensity arguments.  A group (B8) of non-yrast low intensity $\gamma$-ray transitions were found to be in coincidence with the transitions below 31/2$^-$ state of yrast negative parity band B4.  These transitions are placed in the level scheme based on the coincidence and sum energy relationship. The inset of Fig. \ref{sum}  represents  $\gamma-\gamma$ coincidence sum spectrum showing the transitions in group B8.

\section{Discussion}\label{discussion}

\begin{figure}
\begin{center}
\vspace{0.6cm}
\includegraphics[scale=0.44,angle=0]{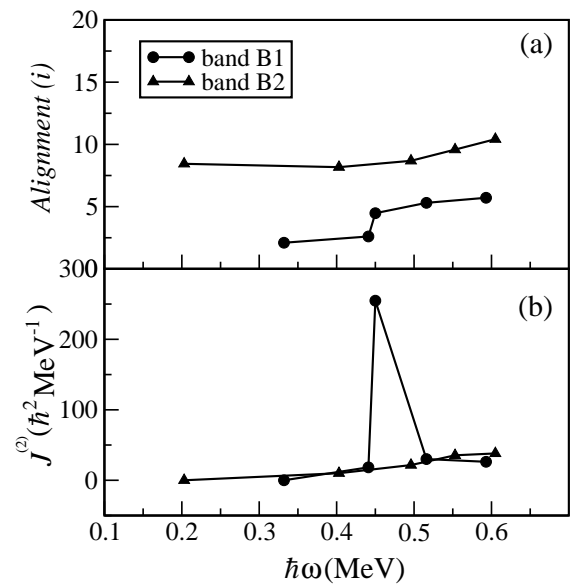}
\caption{\label{xa-dmi}(a) Alignments (upper panel) and (b) dynamic moment of inertia (lower panel) as a function of rotational frequency
for bands B1, B2 of $^{105}$Cd. Harris model parameters of $\Im_0$ = 7.0 $\hbar^2$ MeV$^{-1}$ and $\Im_1$ = 15.0 $\hbar^4$  MeV$^{-3}$ are used in the calculations. }
\end{center}
\end{figure}

\begin{figure}
\begin{center}

\includegraphics[scale=0.44,angle=0]{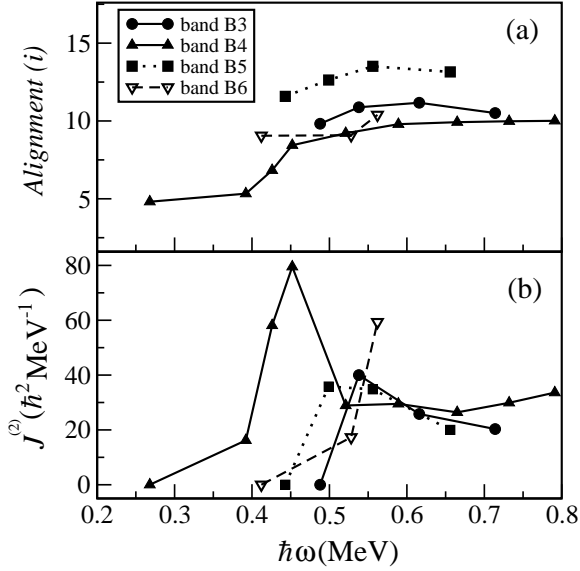}
\caption{\label{xa-dmi-np}(a)Alignments (upper panel) and (b) dynamic moments of inertia (lower panel) as a function of rotational frequency
for bands B3 to B6 of $^{105}$Cd.  Harris model parameters of $\Im_0$ = 7.0 $\hbar^2$ MeV$^{-1}$ and $\Im_1$ = 15.0 $\hbar^4$  MeV$^{-3}$ are used in the calculations. }
\end{center}
\end{figure}

\begin{figure}
\begin{center}

\includegraphics[scale=0.44,angle=0]{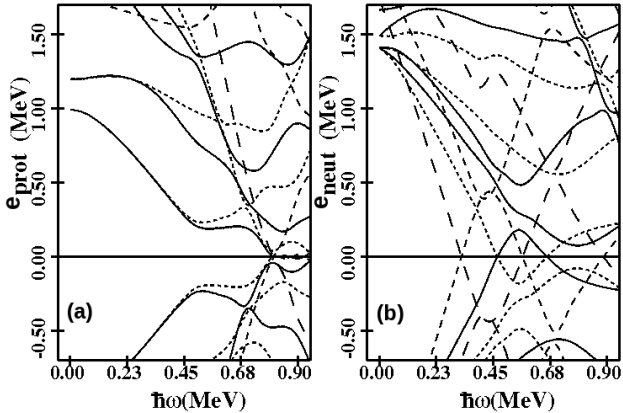}
\caption{\label{QP}The plots show the single quasiparticle routhians based on cranked shell model calculations (a) for protons (left) and (b) for neutrons (right) in  $^{105}$Cd using shape parameters $\beta_2$ = 0.151, and $\gamma \sim$ 6$^\circ$.}
\end{center}
\end{figure}

The structure of $^{105}$Cd was discussed previously by several groups and the possible configurations for the yrast bands have been reported \cite{Regan, D.Jerrestam}. This nucleus lies in a transitional region with two proton holes and seven neutron particles with respect to N = Z = 50 shell closure.  One can, therefore, expect transitional behavior between different nuclear shapes with small or moderate values of quadrupole deformations.  Fig. \ref{compare} shows the negative parity levels based on the $\nu [h_{11/2}]$ band head configuration in the odd-A Cd isotopes \cite{103Cd, 107Cd, 105Cd-AMR, 109Cd}. The energy level spacing for the lowest transition decreasing from 643 keV in $^{103}$Cd to 515 keV in $^{107}$Cd and  522 keV for $^{109}$Cd. This indicates that the odd A $^{103-109}$Cd isotopes shows gradual increase in collectivity with increasing neutron number up to A = 107 and  the collectivity decreases for A =109 (see Fig. \ref{compare}).   This collectivity persists up to spin  27/2 $\hbar$ and 
then 
decreasing gradually for higher spins where the excited band configurations start to dominate leading the nucleus towards non-collective shapes. Such excited bands B3, B5 and B6 are identified in the present work.  

\subsection{Cranked shell model (CSM) analysis}

Band B1 is yrast positive parity band extended to J$^\pi$ = (27/2$^+$)  in this work and band B2 is a newly identified positive parity band. The upper panel of Fig. \ref{xa-dmi} shows the alignments observed in both band B1 and B2.  The quasiparticle alignment has been observed for band B1 around rotational frequency $\approx$ 0.45 MeV, and it can be clearly seen from the lower panel of Fig. \ref{xa-dmi} that the dynamic moment of inertia  peaks at the similar rotational frequency. This alignment observed in band B1 can be interpreted based on the excitations involving  $\nu[ g_{7/2}]$ and  $\nu [d_{5/2}]$ orbitals.  This is also supported by the low alignment gain($\approx$ 3 $\hbar$) observed in band B1.  Band  B2 is depopulated to band B1 through a 1293 keV $\gamma$-transition, which reduces the probability of identification of the band head of B2.   The alignment as well as the dynamic moment of inertia for band B2 in  Fig. \ref{xa-dmi} shows smooth increase with increasing rotational frequency with no 
clear evidence of band crossing though there is slight up bend at $\hbar \omega \sim$ 0.5 MeV.

\begin{figure}
\begin{center}

\includegraphics[scale=0.38,angle=0]{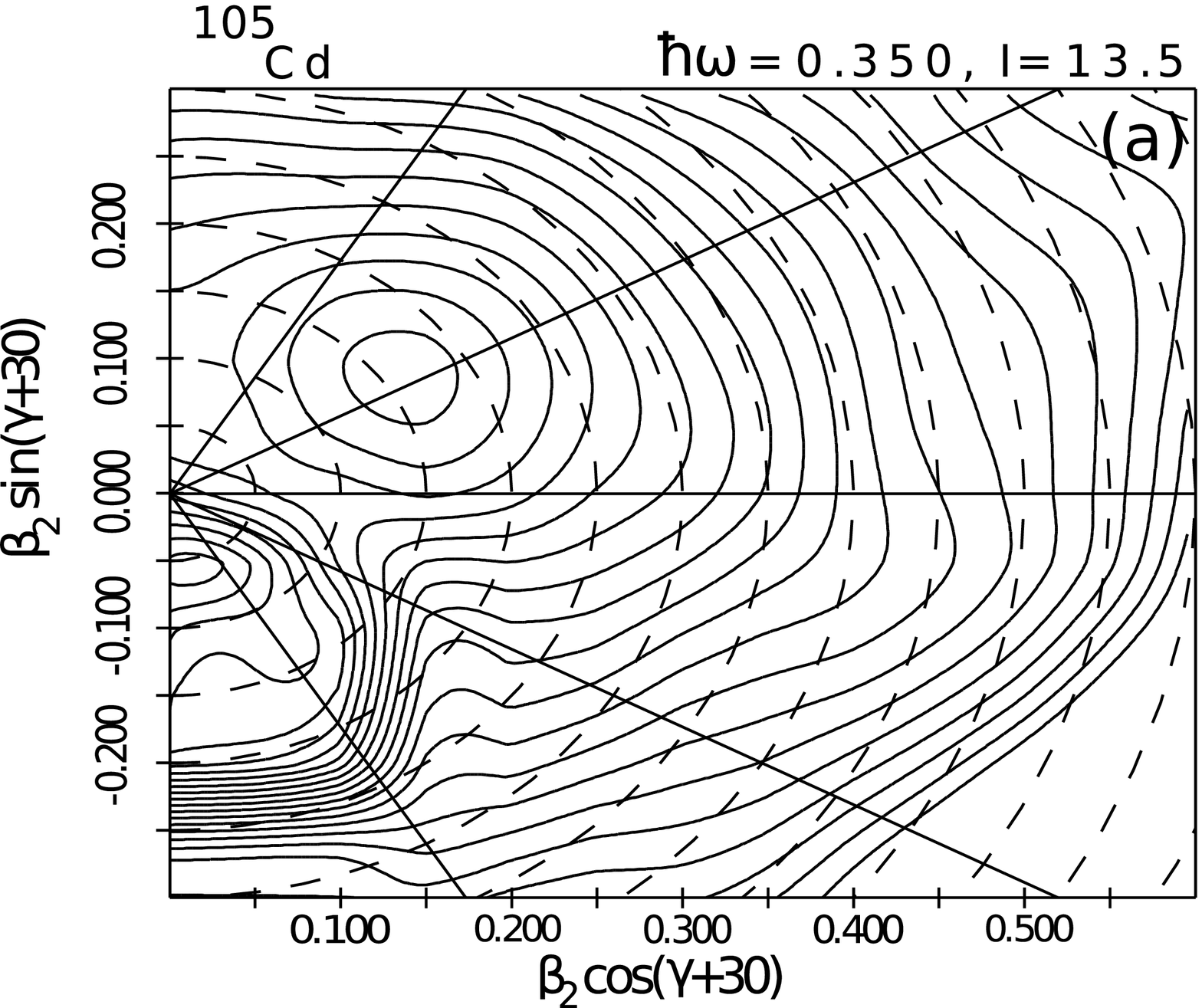}
\includegraphics[scale=0.38,angle=0]{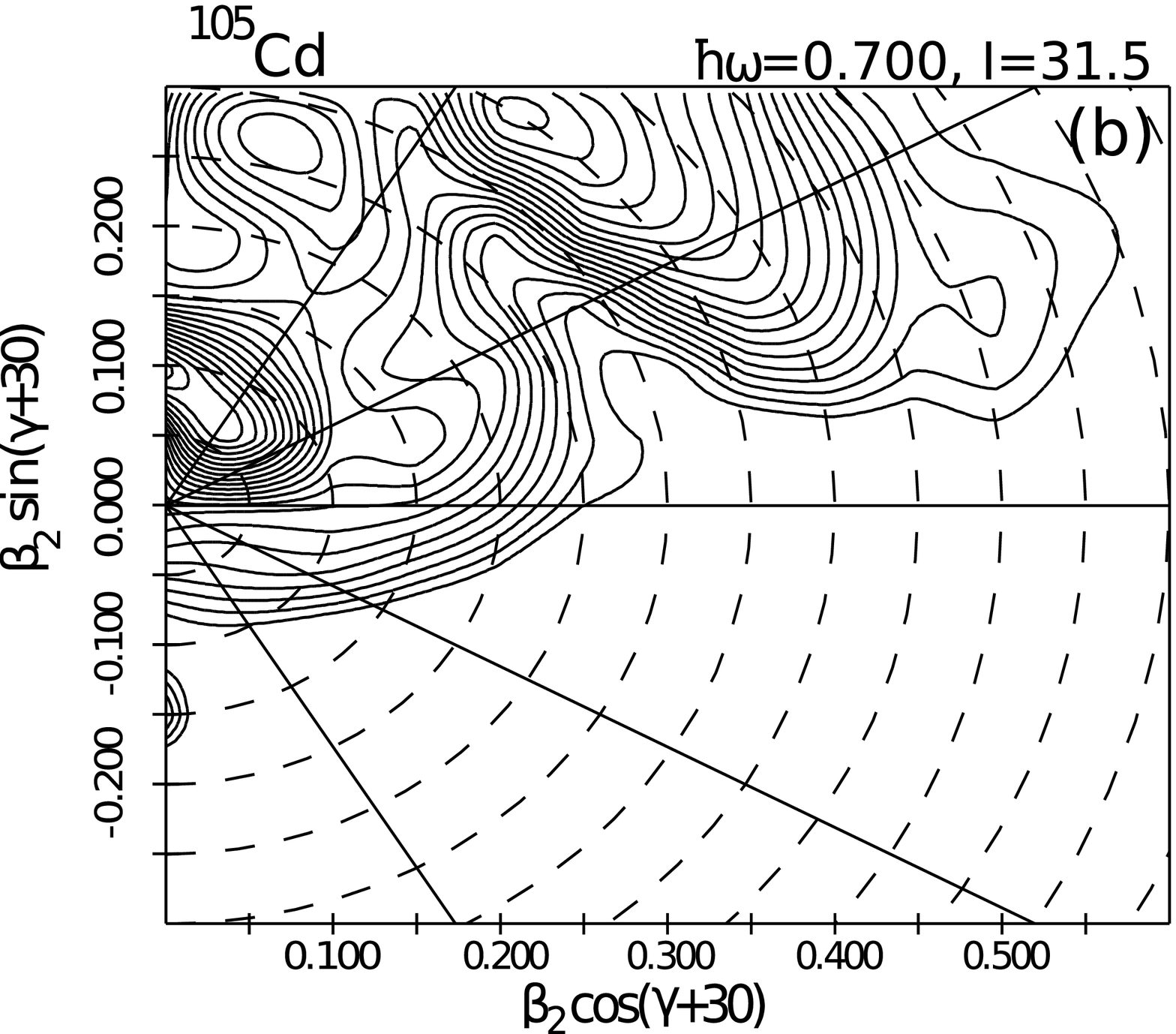}
\caption{\label{TRS}  Total Routhian Surface calculations for the negative parity states ($\pi $, $\alpha $) = (-, -) of $^{105}$Cd  at a rotational frequency of 0.35 MeV (top panel), and 0.70 MeV (bottom panel). The spacing between adjacent contours is 200 keV.}
\end{center}
\end{figure}

Band B4 is strongly populated ground state negative parity yrast band built on  $\nu [h_{11/2}]$ band head configuration.  The upper panel in Fig. \ref{xa-dmi-np} shows the alignments as function of rotational frequency for the negative parity bands. The first band crossing for band B4 has been observed around the rotational frequency $\approx$ 0.44 MeV. This is also evident in the lower panel of Fig. \ref{xa-dmi-np} which shows the dynamic moment of inertia with respect to rotational frequency. The observed alignment in this band is smooth and it is interpreted due to the breaking of $\nu [g_{7/2}]$ pair \cite{D.Jerrestam}.  After the first band crossing, the high spin states in band B4 above spin 27/2 $\hbar$ are generated due to the alignment of a pair of  $ \pi [g_{9/2}]$ proton holes with the configuration  $ \pi [(g_{9/2})^{-2}] \bigotimes \nu [h_{11/2} (g_{7/2})^2)]$. As mentioned earlier the life times of the high spins states above 23/2 $\hbar$ in this band have been reported in Ref. \cite{105Cd-AMR}
 and they have proposed anti-magnetic rotational character to this band after the first band crossing. The alignments for the excited band structures B3, B5 and B6 shows gradual increment with respect to the rotational frequency and the corresponding dynamic moments of inertia (lower panel of Fig. \ref{xa-dmi-np}) shows upbend around rotational frequency $\approx$ 0.54 MeV.  These band structures are primarily based on one and three quasiparticle configurations. 

\begin{figure}
\begin{center}

\includegraphics[scale=0.45,angle=0]{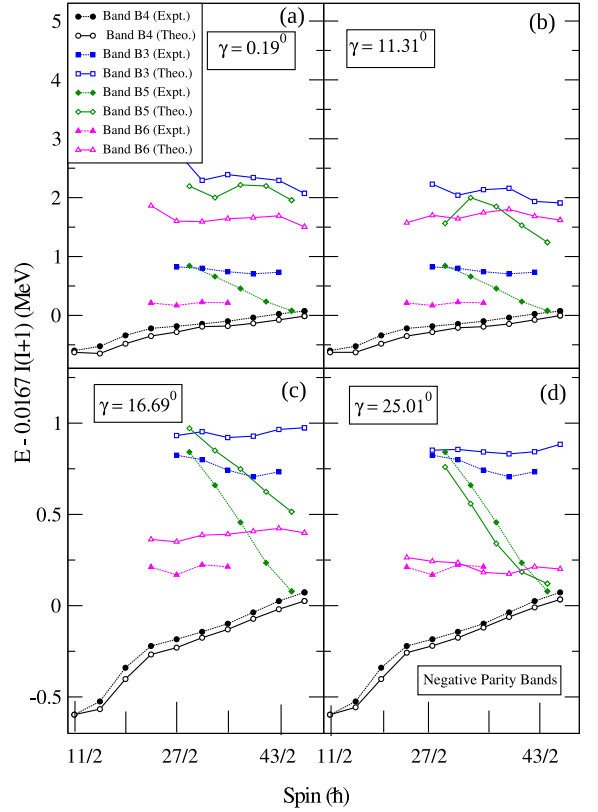}

\caption{(Color online) Comparison of the calculated negative parity band energies minus the liquid drop rotational energy with present experimental data (Band B3, B4, B5 and B6) for $^{105}$Cd. } \label{figtpsm1}
\end{center}
\end{figure}


    To understand the nature of the observed alignments and band crossing frequencies in $^{105}$Cd, at fixed deformation,  single-particle routhians were calculated as a function of rotational frequency  based on a deformed Woods-Saxon potential \cite{HFBB}, including pairing interaction at different shape parameters.  Fig. \ref{QP} shows one of such plots for the quasi particle energies at shape parameters $\beta_2$ = 0.151, $\gamma \approx$ 6$^\circ$ with respect to rotational frequency ($\hbar\omega$). It is evident that the proton crossing frequency is slightly delayed than the neutron crossing frequency which reflects the observed experimental alignments.  Since $^{105}$Cd has odd number of neutrons in g$_{7/2}$, d$_{5/2}$ and h$_{11/2}$ orbitals , the first neutron crossing observed at  $\hbar \omega \approx$ 0.4 MeV is Pauli blocked.  However, the neutron crossing frequency observed  at $\hbar \omega \approx$ 0.45 MeV is a clear indication that the first band crossing in the negative parity band B4 
is due to the alignment of a pair of g$_{7/2}$ neutrons, whereas the proton crossing frequency observed around 0.5 MeV supporting the second alignment is being interpreted as due to the alignment of a pair of g$_{9/2}$ protons.

We have performed Hartree-Fock-Bogoliubov cranking calculations, using the universal parametrization of the Woods-Saxon potential with short range monopole pairing~\cite{HFBB}. BCS formalism was used to calculate the pairing gap $\Delta$ for both protons and neutrons. Total Routhian Surface (TRS) calculations were 
performed in the ($\beta_2$, $\gamma$) plane at different rotational frequencies and the total energy was minimized with respect to hexadecapole 
deformation ($\beta_4$).  The calculations for the negative parity yrast states (-, -) indicate that the nucleus is $\gamma$-soft at low rotational  frequencies ($\hbar\omega$ = 0.1 MeV)  with moderate deformation ($\beta \approx$ 0.13).  The shape of the nucleus changes to prolate ($\beta_2 \approx 0.15, \gamma \approx 6^\circ$) at the rotational frequency $\hbar\omega$ = 0.35 MeV  close to the first band crossing frequency, shown in the top panel of Fig. \ref{TRS}.  At higher rotational frequencies ($\hbar\omega$ = 0.7 MeV), multiple minima are seen in TRS (shown in the bottom panel of Fig. \ref{TRS}) which depicts that the shape of the nucleus is triaxial after the first band crossing.  The validity of the TRS calculations is limited  at higher rotational frequencies, and therefore to understand the triaxial nature of this nucleus and associated band configurations, we have performed triaxial projected shell model calculations described in the following section.


\begin{figure}

\includegraphics[scale=0.45,angle=0]{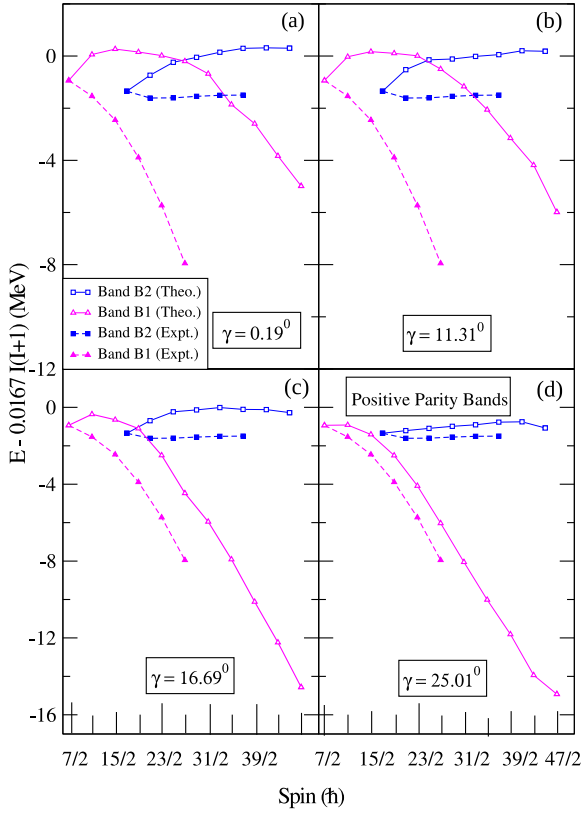}

\caption{(Color online) Comparison of the calculated positive parity band energies minus the liquid drop rotational energy with experimental data (band B1 and B2) for $^{105}$Cd.} \label{figtpsm2}

\end{figure}

\subsection{Triaxial projected shell model (TPSM) analysis}

In order to further analyze the observed band structures, we have also  performed triaxial projected
shell model calculations \cite{JS9,GB12,GB10,JS14,SJ14} for $^{105}$Cd. The advantage of this model as compared to the
CSM approach is that the
angular-momentum is a good quantum number and a direct comparison can be made with experimental data. TPSM calculations are performed in three stages. In the first stage,
triaxial basis are generated by solving the triaxially deformed Nilsson potential with
the deformation parameters of $\beta_2$ and $\gamma$. In the present study of
 $^{105}$Cd, we have used $\beta_2=0.150$ and varied $\gamma$ to investigate the 
triaxial nature of the observed high-spin band-structures. In the second stage, the
intrinsic basis are projected onto good angular-momentum states using the 
three-dimensional angular-momentum projection operator. In the third and final stage,
the projected basis are used to diagonalise the shell model Hamiltonian consisting
of pairing plus quadrupole-quadrupole interaction terms. 

\begin{figure}[htb]
\begin{center}

 \centerline{\includegraphics[trim=0cm 0cm 0cm
0cm,width=0.50\textwidth,clip]{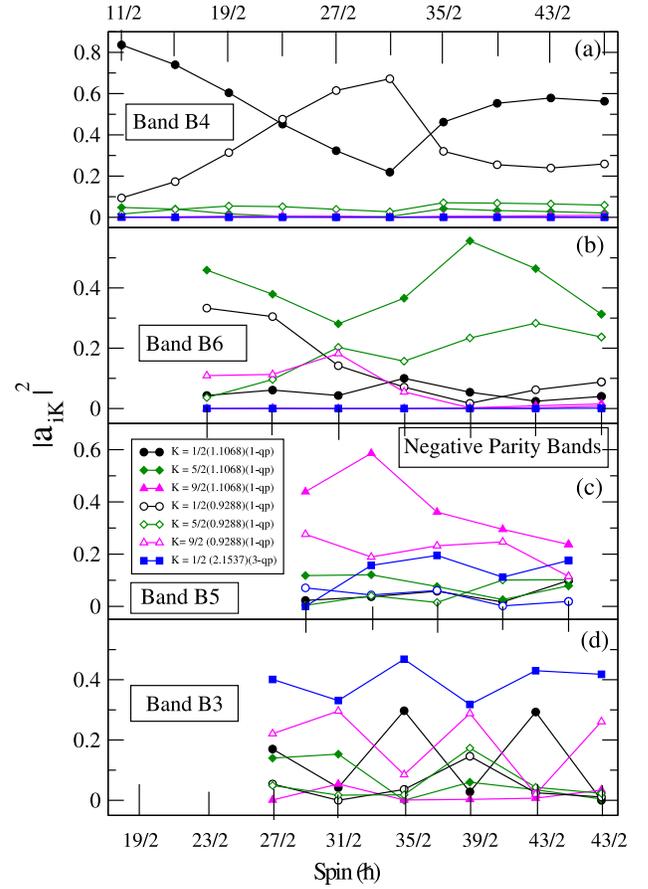}}  

\caption{(Color online)  Probability of various projected K-configurations in the wavefunctions of the observed negative parity bands for $^{105}$Cd (one- and three-quasiparticle
states). 
} \label{figtpsmw1}
\end{center}
\end{figure}

\begin{figure}[htb]
\begin{center}

 \centerline{\includegraphics[trim=0cm 0cm 0cm
0cm,width=0.50\textwidth,clip]{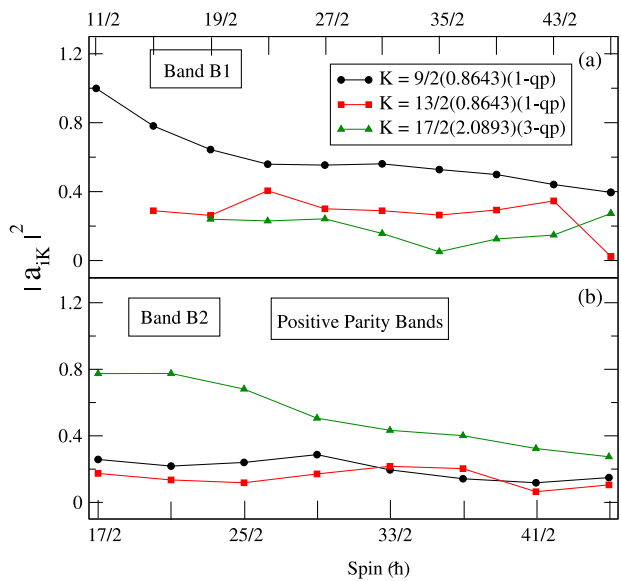}}

\caption{(Color online) Probability of various projected K-configurations in the wavefunctions of the observed positive parity bands for $^{105}$Cd  (one- and three-quasiparticle
states). 
} \label{figtpsmw2}
\end{center}
\end{figure}

The TPSM results obtained after shell model diagonalisation are depicted in Figs. \ref{figtpsm1}
and \ref{figtpsm2} for negative and positive parity bands, respectively.  In Fig. \ref{figtpsm1}, the
results obtained with $\gamma=0.19^0$ which is close to the axial limit, reproduce
the yrast negative-parity band, but the three observed excited bands are totally different
from the TPSM predicted bands. In order to investigate the $\gamma$ dependence of the
observed bands, we have also performed TPSM study for $\gamma=11.31^0, 16.69^0$ and $25.01^0$. The
results depicted in Fig. \ref{figtpsm1} reveal that $\gamma=25.01^0$ reproduces the observed 
bands quite reasonably. It is noted from the figure that the yrast negative-parity band is
almost unchanged with $\gamma$ and, therefore, indicating that the yrast band 
is $\gamma$-soft. However, the excited bands are noted to have a strong dependence on
$\gamma$ and are predicted to have $\gamma$-rigid deformation. We would like to mention that TPSM
calculations were carried out in terms of the deformation parameters, $\epsilon$ and $\epsilon'$,
and when expressing them in terms of $\gamma$ results into fractional values for $\gamma$.
 
The result for the lowest two positive-parity bands are shown in Fig. \ref{figtpsm2} 
and are again predicted to have a strong dependence on the triaxial deformation. The
observed positive parity bands are noted to be reproduced for $\gamma=25.01^0$
and therefore it is predicted from the present investigation that the shape of
$^{105}$Cd changes from $\gamma$-soft to $\gamma$-rigid for excited configurations.

In order to shed light on the structure of the observed bands, the wavefunctions 
of the bands are plotted in Figs.~\ref{figtpsmw1} and \ref{figtpsmw2}. The band B4
shown in Fig.~\ref{figtpsmw1}(a) is predominantly composed of one-quasineutron configuration,
K=1/2(1.1068) an K=1/2(0.9288) in the low-spin region.  The bands B6 and  B5
are predominantly composed of the one-quasineutron configurations, K=5/2 (1.1068) and K=5/2(0.9288) and K= 9/2(1.1068) and K= 9/2(0.9288) and the band 
B3 is primarily a three-quasiparticle configuration. It is evident from Fig.~\ref{figtpsmw2}
that two observed positive parity B1 and B2 bands have predominant one-quasiparticle 
configurations, K=9/2, 13/2 and also three-quasiparticle contribution with K=17/2.  We would like to point out that TPSM results in Figs.~\ref{figtpsmw1} and \ref{figtpsmw2}  deviate  for higher spin states, in particular, for bands B3 and B2. This indicates  the importance of five or more quasiparticles states which are not included in the present TPSM analysis. We are presently in the process of generalizing the TPSM approach to include higher quasiparticle states. 


\section{Summary}\label{summary}
In the present work, high spin states in $^{105}$Cd have been studied using $^{92}$Mo($^{16}$O, 2pn)$^{105}$Cd reaction at an incident beam energy of 75 MeV. The earlier reported level structures have been verified. We have incorporated several modifications in the positive and negative parity bands. Four new band structures have been established with the addition of 30 new transitions to the level scheme.  Spin and parity of  the observed states have been assigned based on the directional correlation orientation ratios, polarization asymmetry values and  systematics.  The experimental alignments and rotational properties of the identified bands are discussed within the framework of cranked shell model, which explains the observed experimental alignments reasonably well.  The TRS calculations for the yrast band structures show evolution from  prolate shape at low rotational frequency to triaxial shapes at higher rotational frequencies. The triaxial projected shell model calculations are performed, which predict that, the observed band 
structures has strong dependence on triaxial deformation and indicate  that the shape of $^{105}$Cd changes from $\gamma$-soft to $\gamma$-rigid for excited configurations.  The calculated wave functions for the newly observed excited bands predict that one-quasiparticle configurations with K = 5/2 and 9/2 components are dominant for negative parity bands B5 and B6. The positive parity band B2 is based on configurations  with K = 9/2, 13/2 (one-quasiparticle)  and also K=17/2 (three-quasiparticle).

\section{Acknowledgement}
The authors would like to thank target laboratory group, IUAC for their help in making $^{92}$Mo target.  We thank Prof. S. C. Pancholi for helpful discussions and comments. The TIFR pelletron crew and INGA collaboration are also acknowledged for making this experiment possible.  The financial support provided by DST for INGA project (No.IR/S2/PF-03/2003-1) is gratefully acknowledged.


\begin{thebibliography}{100}

\bibitem{103Cd} A. Chakraborty, Krishichayan, S. Mukhopadhyay, S. Ray, S. N. Chintalapudi, S. S. Ghugre, N. S. Pattabiraman, A. K. Sinha, S. Sarkar, U. Garg, S. Zhu, and M. Saha Sarkar, Phys. Rev. C {\bf 76}, 044327 (2007).

\bibitem{104Cd} S. D. Robinson, S. J. Freeman, D. P. Balamuth, M. Carpenter, M. Devlin, B. G. Dong, J. L. D\"{u}rell, P. Hausladen, D. R. LaFosse, T. Lauritsen, M. J. Leddy, I. Y. Lee, R. McLeod, C. J. Lister, A. O. Macchiavelli, I. Ragnarsson, D. G. Sarantities, D. Seweryniak, R. B. E. Taylor, and B. J. Varley, J. Phys. G {\bf 28}, 1415 (2002).

\bibitem{mag} S. H. Yao, H. L. Ma, L. H. Zhu, X. G. Wu, C. Y. He, Y. Zheng, B. Zhang, G. S. Li, C. B. Li, S. P. Hu, X. P. Cao, B. B. Yu, C. Xu, and Y. Y. Cheng,  Phys. Rev. C {\bf 89}, 014327 (2014).

\bibitem{anti} S. Zhu, U. Garg, A. V. Afanasjev, S. Frauendorf, B. Kharraja, S. S. Ghugre, S. N. Chintalapudi, R. V. F. Janssens, M. P. Carpenter, F. G. Kondev, and T. Lauritsen, Phys. Rev. C {\bf 64}, 041302(R) (2001). 

\bibitem{vib-rot} P. H. Regan, C. W. Beausang, N. V. Zamfir, R. F. Casten, Jing-ye Zhang, A. D. Yamamoto, M. A. Caprio, G. G\"{u}rdal, A. A. Hecht, C. Hutter, R. Kr\"{u}cken, S. D. Langdown, D. A. Meyer, and J. J. Ressler, Phys. Rev. Lett. {\bf 90}, 152502 (2003).

\bibitem{beta12} S. F. Ashley, P. H. Regan, K. Andgren, E. A. McCutchan, N. V. Zamfir, L. Amon, R. B. Cakirli, R. F. Casten, R. M. Clark, W. Gelletly, G. G\"{u}rdal, K. L. Keyes, D. A. Meyer, M. N. Erduran, A. Papenberg, N. Pietralla, C. Plettner, G. Rainovski, R. V. Ribas, N. J. Thomas, J. Vinson, D. D. Warner, V. Werner, E. Williams, H. L. Liu, and F. R. Xu, Phys. Rev. C {\bf 76}, 064302 (2007).

\bibitem{beta22} R. F. Casten, N. V. Zamfir, and D. S. Brenner, Phys. Rev. Lett. {\bf 71}, 227 (1993).

\bibitem{D.Jerrestam}D. Jerrestam, B. Fogelberg,  A. Kerek, W. Klamra, F. Lid\'{e}n, L. O. Norlin, J. Kownacki, D. Seweryniak, Z. \.{Z}elazny, C. Fahlander,  J. Nyberg, M. Guttormsen, J. Rekstad, T. Spedstad-Tveter, A. Gizon, J. Gizon, R. Bark, G. Sletteng, M. Piiparinen, Z. Preibisz, T.E Thorsteinsen, E. Ideguchi, and S. Mitarai,  Nucl. Phys. A{\bf 593},162 (1995).

\bibitem{107Cd} Dan Jerrestam, F. Lid\'{e}n, J. Gizon, L. Hildingsson, W. Klamra, R. Wyss, D. Barn\'{e}oud, J. Kownacki, Th. Lindblad, and J. Nyberg, Nucl. Phys. A {\bf 545}, 835 (1992).


\bibitem{alpha} J. Genevey-Rivier, J. Trrherne, J. Danirre, R. Brraud, M. Meyer, and R. Rougny, J. Phys. G {\bf 4}, 953 (1978).

\bibitem{16O} D. C. Stromswold, D.O. Elliott, Y. K. Lee, L. E. Samuelson, J. A. Grau, F. A. Rickey and P. C. Simms, Phys. Rev. C {\bf 17}, 143 (1978).


\bibitem{Regan}P. H. Regan, G. D. Dracoulis, G. J. Lane,  P. M. Walked, S. S. Anderssen, A. P. Byme, P. M. Davidson, T. KibMi, A. E. Stuchbew, and K. C. Yeungt,  J. Phys. G {\bf 19}, L157 (1993).

\bibitem{105Cd-AMR} Deepika Choudhury, A. K. Jain, M. Patial, N. Gupta,  P. Arumugam,  A. Dhal,  R. K. Sinha,  L. Chaturvedi, P. K. Joshi, T. Trivedi,  R. Palit,  S. Kumar,  R. Garg, S. Mandal,  D. Negi, G. Mohanto, S. Muralithar, R. P. Singh, N. Madhavan,  R. K. Bhowmik, and S. C. Pancholi, Phys. Rev. C {\bf 82}, 061308(R) (2010).

\bibitem{DAE} M. Kumar Raju, D. Negi, S. Muralithar, R. P. Singh, R. Kumar, Indu Bala, T. Trivedi, A. Dhal, K. Rani, R. Gurjar, D. Singh, J. Kaur, R. Palit, B. S. Naidu, S. Saha, J. Sethi, and R. Donthi,  Proceedings of the DAE Symp. on Nucl. Phys. {\bf 58} (2013).

\bibitem{INGA}R. Palit, AIP Conf. Proc. {\bf 1336}, 573 (2011).

\bibitem{pace} A.Gavron, Phys.Rev. C {\bf 21}, 230 (1980).

\bibitem{DDAQ}R. Palit, S. Saha, J. Sethi, T. Trivedi, S. Sharma, B. S. Naidu, S. Jadhav, R. Donthi, P. B. Chavan, H. Tan, and W. Hennig, Nucl. Instrum. Methods Phys. Res. A {\bf 680}, 90 (2012).

\bibitem{radware} D. C. Radford, Nucl. Instrum. Methods Phys. Res. A {\bf 361}, 297 (1995).


\bibitem{candle}B. P. Ajith Kumar, E. T. Subramaniam, K. M. Jayan, S. Mukherjee, and R. K. Bhowmik, in Proceeding on Symp. on
Adv. in Nucl. and Allied Instrum. India, 51 (1997).

\bibitem{dco} A. Kr\"{a}mer-Flecken, T. Morek, R. M. Lieder, W. Gast, G. Hebbinghaus, H. M. J\"{a}ger, and W. Urban, Nucl. Instrum. Methods Phys. Res. A {\bf 275}, 333 (1989).

\bibitem{pol1} G. Duchene et al., Nucl. Instrum. Methods Phys. Res. A {\bf 432}, 90 (1999).

\bibitem{pol2} K. Starosta et al., Nucl. Instrum. Methods Phys. Res. A {\bf 423}, 16 (1999).

\bibitem{105In} D. Kast, A. Jungclaus, K.P. Lieb, M. G\,{o}rska, G. de Angelis, P.G. Bizzeti, A. Dewald, C. Fahlander, H. Grawe, R. Peusquens, M. De Poli, H. Tiesler, Eur. Phys. J. A 3, 115-128 (1998).

\bibitem{isomer} D. De Frenne, E. Jacobs, M. Verboven, and P. Gelder, Nucl. Data Sheets {\bf 47}, 261 (1986).

\bibitem{103Pd} B. M. Nyak\'{o}, J. Gizon, A. Gizon, J. Tim\'{a}r, L. Zolnai, A. J. Boston, D. T. Joss, E. S. Paul, A. T. Semple, N. J. \,{O}Brien, C. M. Parry, A. V. Afanasjev, and I. Ragnarsson, Phys. Rev. C {\bf 60}, 024307 (1999).

\bibitem{101Ru}A. D. Yamamoto, P. H. Regan, C. W. Beausang, F. R. Xu, M. A. Caprio, R. F. Casten, G. G\"{u}rdal, A. A. Hecht, C. Hutter, R. Kr\"{u}cken, S. D. Langdown, D. Meyer, J. J. Ressler, and N. V. Zamfir, Phys. Rev. C {\bf 66}, 024302 (2002).


\bibitem{109Cd} C. J. Chiara, S. J. Asztalos, B. Busse, R. M. Clark, M. Cromaz, M. A. Deleplanque, R. M. Diamond, P. Fallon, D. B. Fossan, D. G. Jenkins, S. Juutinen, N. S. Kelsall, R. Kr\"{u}cken, G. J. Lane, I. Y. Lee, A. O. Macchiavelli, R. W. MacLeod, G. Schmid, J. M. Sears, J. F. Smith, F. S. Stephens, K. Vetter, R. Wadsworth, and
S. Frauendorf, Phys. Rev. C {\bf 61}, 034318 (2000).

\bibitem{harris} S. M. Harris, Phys. Rev. {\bf 138}, B509 (1965).

\bibitem{HFBB} W. Nazarewicz, J. Dudek, R. Bengtsson, T. Bengtsson, and I. Ragnarsson, Nucl. Phys. A {\bf 435}, 397 (1985).
\bibitem{JS9} J. A. Sheikh and K. Hara, Phys. Rev. Lett. {\bf 82}, 3968 (1999).
\bibitem{GB12} G. H. Bhat, J. A. Sheikh, and R. Palit, Phys. Lett. B {\bf  707}, 205 (2012).
\bibitem{GB10} J. A. Sheikh, G. H. Bhat, Y. Sun, and R. Palit, Phys. Lett. B {\bf  688},   305  (2010).
\bibitem{JS14} G. H. Bhat, W. A. Dar, J. A. Sheikh, and Y. Sun, Phys. Rev. C {\bf  89}, 014328 (2014).
\bibitem{SJ14} G. H. Bhat, R. N. Ali, J. A. Sheikh, and R. Palit, 
               Nucl. Phys. A {\bf  922}, 150 (2014).
\end{thebibliography}
\end{document}